\documentclass[aps,prb,twocolumn,showpacs,amsmath,amssymb,superscriptaddress,longbibliography,floatfix]{revtex4-1}

\usepackage{amsmath, amsfonts, amssymb, bm}
\usepackage{float}
\usepackage{mathtools}
\usepackage[inline]{enumitem}
\usepackage{graphicx}
\usepackage{xcolor}
\usepackage{lineno}
\usepackage[breaklinks,hypertexnames=false]{hyperref}
\hypersetup{colorlinks,linkcolor={magenta},citecolor={blue},urlcolor={blue}}
\usepackage[capitalize]{cleveref}

\usepackage{glossaries}
\glsdisablehyper    
\newacronym{qsh}{QSHI}{quantum spin Hall insulator}
\newacronym{mkp}{MKP}{Majorana Kramers pair}
\newacronym{sgs}{SGS}{subharmonic gap structure}
\newacronym{mar}{MAR}{multiple Andreev reflections}

\usepackage[utf8x]{inputenc}
\usepackage{amssymb}
\usepackage[caption=false]{subfig}
\usepackage{pstricks}
\usepackage{relsize}
\usepackage{bm}
\usepackage{braket}
\usepackage{epstopdf}

\usepackage[all]{hypcap}

\begin{document}
\title{Phase-tunable multiple Andreev reflections in a quantum spin Hall strip}

\newcommand{\tianjin}{Department of Physics, Tianjin University, Tianjin 300072, China}

\newcommand{\madrid}{Department of Theoretical Condensed Matter Physics, Condensed Matter Physics Center (IFIMAC) and Instituto Nicol\'as Cabrera, Universidad Aut\'onoma de Madrid, 28049 Madrid, Spain}

\author{Xue Yang}
\affiliation{\tianjin}

\author{Pablo Burset}
\affiliation{\madrid}

\author{Bo Lu}
\affiliation{\tianjin}

\date{\today}

\begin{abstract}
   A quantum spin Hall strip where different edges are contacted by $s$-wave superconductors with a phase difference $\phi$ supports Majorana bound states protected by time-reversal symmetry. 
   We study signatures of these states in a four-terminal setup where two Josephson junctions are built on opposite edges of the strip and the phase difference between superconductors can be controlled by an external flux. Applying a voltage bias across the quantum spin Hall strip results in a sequence of conductance peaks from multiple Andreev reflections. 
   We find that this so-called subharmonic gap structure is very sensitive to the phase difference and displays a phase-controlled even-odd effect, where all odd spikes disappear when the Majorana states are formed for $\phi=\pi$. Moreover, the remaining even spikes split when the superconductors forming the junction have different gap size. 
   We explain these features by showing that any midgap bound states enhance the transmission of the even order multiple Andreev reflections, while the reduced density of states at the gap edges suppresses the odd order ones. 
\end{abstract}
\maketitle

\section{Introduction.}

The \gls{qsh}~\cite{Kane051,Kane052,Bernevig06,Liu08,Wu06,Xu06} is a prominent topological material that is recently attracting significant attention. 
Its defining feature is the emergence of helical, or spin-momentum locked, edge states where different spins circulate in opposite directions. 
These helical edge states have been measured in experiments~\cite{Roth09,Brune12} and provide a pathway to develop novel quantum phenomena and functionalities~\cite{Hsu_2021}. 
For example, \glspl{qsh} are predicted to host Majorana bound states with revolutionary prospects in fault-tolerant quantum computations~\cite{Kitaev01,Kitaev03,Nayak08,Hasan10,XLQi11,Alicea11,Leijnse12,Ando13,Tkachov13,Kitaev01}. 
Such topological superconductivity can be generated in helical states with or without time-reversal invariance.  
Breaking time-reversal symmetry, a single helical edge can be proximitized by ferromagnets and superconductors so that Majorana modes appear at the boundaries between them~\cite{Kwon04,FuL09,Fu09}. 
However, combining ferromagnets with superconductors in \glspl{qsh} is experimentally challenging due to the detrimental effects of the magnetic exchange on the proximity-induced gap. Therefore, efforts are devoted to propose platforms without magnetic materials that realize time-reversal invariant topological superconductors with Kramers pairs of zero-energy Majorana bound states. 
These so-termed \glspl{mkp} are twofold degenerate~\footnotemark[1], leading to a quantized conductance of $4e^2/h$~\cite{Kim16} and mirror fractional Josephson effect~\cite{ZhangFan13} as experimental signatures. 

\footnotetext[1]{Kramers theorem states that, in a time-reversal symmetric system with half-integer total spin, for any energy eigenstate its time-reversal state is also an eigenstate with the same energy. }

\begin{figure}[hb]
	\includegraphics[width=0.90\columnwidth]{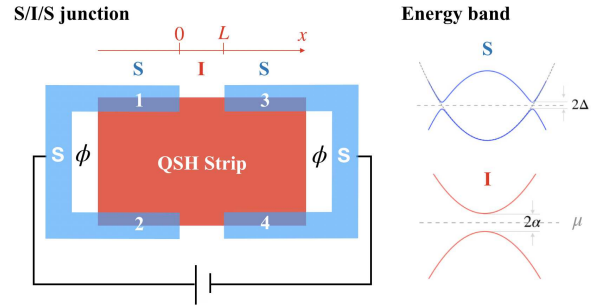}
	\caption{
		Schematic diagram of the device layout to detect a topological transition. Terminals 1 and 2 (3 and 4) connect to the left (right) superconducting electrodes and the superconducting phase difference between terminals is controlled by a magnetic flux. The band structure of low-energy quasiparticles near the Fermi surface in each region is shown below. 
	}
	\label{Fig1}
\end{figure}

In one approach to realize \glspl{mkp}, the two opposite edges of a \gls{qsh} strip are coupled to superconducting leads with a phase difference of $\pi $~\cite{Berg13,Roman16,Blasi19}. An experimental signature to detect \glspl{mkp} in such a Josephson junction is the \gls{sgs}; a series of resonant conductance peaks in a voltage-biased Josephson junction. 
The \gls{sgs} is generated by \gls{mar} when two superconductors are in electric contact and a voltage bias drives sequential Andreev reflections of quasiparticles at the interface between them. Incident quasiparticles gain or lose an energy $eV$ as they travel across the interface, until escaping to the reservoirs for energies above the superconducting gap. In conventional Josephson junctions, the peaks in the \gls{sgs} come from the singular density of states at the superconducting energy gap edges hosting incident and escaping quasiparticles~\cite{Bo20}. The conductance peaks are thus positioned at $eV_{n}\!=\!2\Delta_0 /n$, with $\Delta_0$ the superconducting gap and $n$ an integer number~\cite{KLA82,OTBK83,Arnold87,Bratus95,Averin95,Cuevas96}. By contrast, due to the presence of Majorana modes in topological Josephson junctions, new resonant channels form in the middle of the energy gap leading to an anomalous \gls{sgs} with only even integer peaks. That is, the conductance resonances are located at $eV_{m}=2\Delta_{0}/m$ where $m$ is now an even integer~\cite{Meyer11,San_Jose_2013,Zazu16,Setiawan172}. 
Anomalous \glspl{sgs} have been predicted in Josephson junctions mediated by the edge states of a \gls{qsh}, which were interpreted as a parity-changing process in a topological Josephson junction~\cite{Meyer11}. 
Angle-resolved \gls{sgs} have also been explored in two-dimensional Josephson junctions for detecting chiral Majorana states~\cite{Linde23}.
However, such setups require breaking time-reversal symmetry by Zeeman coupling from a magnetic material at the \gls{qsh} edges, thus hindering their possible experimental implementation. 

In this paper we consider a time-reversal invariant \gls{qsh} strip with no magnetic elements to simplify these experimental challenges in the search for Majorana states. In our approach, two opposite strip edges are covered by superconductors with a tunable phase difference, see \cref{Fig1}. 
Importantly, the strip width is such that the edge states are not decoupled. 
Consequently, the edge states hybridize and their characteristic Dirac-like linear dispersion becomes gapped by $~2\alpha$, with $\alpha$ the inter-edge coupling strength. 
The inter-edge coupling opens a reflection channel between particles at opposite edges, without breaking time-reversal symmetry. 
The resulting tunnel Josephson junction has variable transmission controlled by $\alpha$. 
We then consider a bias voltage at terminals 1 and 2 that injects a current collected by terminals 3 and 4 (\cref{Fig1}). At the same time, the phase difference $\phi$ between superconductors can be controlled by a magnetic flux. For $\phi\!=\pi $ the system preserves time-reversal symmetry and hosts \glspl{mkp}. 

Interestingly, the \gls{sgs} for this junction can be tuned by $\phi $ in stark contrast to conventional Josephson junctions. 
For $\phi=0$, the junction behavior is in good agreement with conventional BCS Josephson junctions~\cite{1997ac}: 
the \gls{sgs} features peaks at $eV_m=2\Delta_{L/R}/m$ and $eV_n=(\Delta_L+\Delta_R)/n$, with $m$ and $n$ being, respectively, even and odd integers, and $\Delta_L$ ($\Delta_R$) the left (right) pair potential. As a result, only the even spikes of the \gls{sgs} split for asymmetric junctions with $\Delta_L \neq \Delta_R$. 
By contrast, when time-reversal invariant \glspl{mkp} emerge for $\phi=\pi$, the junction behavior is very anomalous: 
only the even resonant peaks ($eV_m=2\Delta_{L/R}/m$) appear, and all spikes split for asymmetric junctions. 
We explain this anomalous behavior by showing that midgap bound states from \glspl{mkp} are only connected to even order \gls{mar}, while odd order processes are only sensitive to the gap edges. Since the emergence of \glspl{mkp} is associated with an enhanced density of states at zero energy and a reduction of it at the gap edges, only even \gls{mar} survive for $\phi=\pi$, while both even and odd conductance peaks appear otherwise. 
%
The proposed time-reversal invariant multi-terminal Josephson junction can thus help circumvent some of the experimental challenges in the search for Majorana bound states on \gls{qsh}-based superconducting heterostructures. 

The rest of the paper is organized as follows. In \cref{sec:model}, we describe the model. 
We present the transport properties of symmetric and asymmetric junctions in \cref{sec:SIS} and \cref{sec:SIS-2}, respectively. 
Finally, we conclude this work with a brief summary in \cref{sec:conc}. 
We also present the tunneling conductance of a normal-superconductor junction and further details of our calculations in the Appendix.

\section{Model and Formalism. \label{sec:model}}

The low-energy effective edge state Hamiltonian is given by $\mathcal{H=H}%
_{0}+\mathcal{H}_{S}$ with
\begin{align}
	\mathcal{H}_{0}={}& \int dx\hat{\Psi}^{\dag }\left[ -i\hbar v\hat{\sigma}_{z}%
	\hat{s}_{z}\partial _{x}+\alpha \hat{\sigma}_{x}-\mu \right] \hat{\Psi}, \\
	\mathcal{H}_{S}={}& \Delta \int dx\left[ \hat{\Psi}_{1,\uparrow }^{\dag }%
	\hat{\Psi}_{1,\downarrow }^{\dag }+e^{i\phi }\hat{\Psi}_{-1,\uparrow }^{\dag
	}\hat{\Psi}_{-1,\downarrow }^{\dag }\right] +\mathrm{h.c.},
\end{align}%
and basis $\hat{\Psi}=(\hat{\Psi}_{1,\uparrow }^{\dag },\hat{\Psi}%
_{1,\downarrow }^{\dag },\hat{\Psi}_{-1,\uparrow }^{\dag },\hat{\Psi}%
_{-1,\downarrow }^{\dag })^{T}.$ Here, the subscript $\sigma \in \{1,-1\}$
labels the different edges and the Pauli matrices $\hat{\sigma}_{i}$ and $%
\hat{s}_{i}$, with $i\in \{x,y,z\}$, act on edge and spin spaces,
respectively. The chemical potential is $\mu \left( x\right) =\mu \left[
\Theta \left( -x\right) +\Theta \left( x-L\right) \right] $, with $\Theta
\left( x\right) $ being the Heaviside step function and $L$ the junction
length, and $\alpha $ is the coupling strength between opposite edges~\cite%
{Niu08}. 
We further assume that the chemical potential for the insulating region $(0<x<L)$ is tuned to the middle of the gap, but remains large ($\mu \gg \alpha$) for the S regions, thus forming a tunnel junction of variable transmission $D=tt^*$ with $t=1/\cosh \left( \alpha L/v\right) $. It can be seen that $D$ is tunable by changing the junction length $L$ with a fixed $\alpha$.
We define the pair potential $\Delta \left( x\right) =\Delta _{L}\Theta \left( -x\right) +\Delta_{R}\Theta \left( x-L\right) $.
Using the Bogoliubov transformation $\hat{\Psi}_{\sigma
	,s}=\sum_{N}u_{\sigma ,s}^{N}\hat{\gamma}_{N}+v_{\sigma ,s}^{N,\ast }\hat{%
	\gamma}_{N}^{\dag }$, we derive the Bogoliubov-de Gennes (BdG) Hamiltonian $H=H_{1}\oplus H_{2}$, with
\begin{equation}
	H_{\eta =1,2}=%
	\begin{pmatrix}
		\hat{h}_{i} & \hat{\Delta}_{i} \\
		\hat{\Delta}_{i}^{\dag } & -\hat{h}_{i}%
	\end{pmatrix} ,
\end{equation}
and
\begin{align}
	\hat{h}_{1\left( 2\right) } ={}&
	\begin{pmatrix}
		\mp i\hbar v\partial _{x}-\mu  & \alpha  \\
		\alpha  & \pm i\hbar v\partial _{x}-\mu
	\end{pmatrix}
	, \\
	\hat{\Delta}_{1\left( 2\right) } ={}&
	\begin{pmatrix}
		\pm \Delta  & 0 \\
		0 & \pm \Delta e^{i\phi }%
	\end{pmatrix}	,
\end{align}
where $H_{1}$ ($H_{2}$) acts on $(u_{1,\uparrow }^{N},u_{-1,\uparrow
}^{N},v_{1,\downarrow }^{N},v_{-1,\downarrow }^{N})^{T}$ [$(u_{1,\downarrow
}^{N},u_{-1,\downarrow }^{N},v_{1,\uparrow }^{N},v_{-1,\uparrow }^{N})^{T}$]
space. The wavefunctions can be found in Appendix~\ref{AppA}.

The time-dependent wavefunctions at the central scattering region, (0, $L$), for an incident quasiparticle from terminal 1 read
\begin{equation}
	\Phi _{x=0^{-}}=\sum_{n}e^{-i\frac{\left( \epsilon +2neV\right) t}{\hbar }}%
	\begin{pmatrix}
		{}(J_{\epsilon }^{\left( 1\right) }\delta _{n,0}+a_{L,2n}A_{n}) \\
		B_{n} \\
		A_{n} \\
		a_{L,2n}e^{-i\phi }B_{n}%
	\end{pmatrix}%
	,  \label{Wave1}
\end{equation}%
\begin{equation}
	\Phi _{x=L^{+}}=\sum_{n}e^{-i\frac{\left[ \epsilon +\left( 2n+1\right) eV%
			\right] t}{\hbar }}%
	\begin{pmatrix}
		C_{n} \\
		a_{R,2n+1}e^{i\phi }D_{n} \\
		a_{R,2n+1}C_{n} \\
		D_{n}%
	\end{pmatrix},
\end{equation}
where $a_{L/R,n}\equiv a_{L/R}\left( \epsilon +neV\right) $ with
\begin{equation}
a_{L/R}\left( \epsilon \right) =\left\{
\begin{array}{lc}
\frac{\epsilon -\text{sgn}\left( \epsilon \right) \left( \epsilon
	^{2}-\Delta _{L/R}^{2}\right) ^{1/2}}{\Delta _{L/R}}, & \left\vert \epsilon
\right\vert >\Delta _{L/R} \\
\frac{\epsilon -i\left( \Delta _{L/R}^{2}-\epsilon ^{2}\right) ^{1/2}}{%
	\Delta _{L/R}}, & \left\vert \epsilon \right\vert <\Delta _{L/R}.%
\end{array} ,
\right.
\end{equation}%
and $J_{\epsilon }^{\left( 1\right) }=\sqrt{1-\left\vert a_{L} ( \epsilon ) \right\vert ^{2}}$ being the amplitude of the incident quasiparticle from terminal 1 into the scattering region. 
The wavefunctions $\Phi _{0^{-}}$ and $\Phi_{L^{+}}$ are connected by the scattering matrices
\begin{equation}
S_{e}=S_{h}^{\ast }=\left[
\begin{array}{cc}
r & t \\
t & -r^{\ast }t/t^{\ast }%
\end{array}%
\right] ,
\end{equation}%
with $r=-i\tanh \left( \alpha L/v\right) $. 
Here, we have assumed that $\alpha \gg \Delta _{L/R}$ and thus the scattering matrices can be approximated as
energy independent.  
Consequently, the coefficients $A_{n}$, $B_{n}$, $C_{n}$, and $D_{n}$ are related by
\begin{align}
\begin{pmatrix}
B_{n} \\
C_{n}%
\end{pmatrix}
= {}& S_{e}
\begin{pmatrix}
J_{\epsilon }^{\left( 1\right) }\delta _{n,0}+a_{L,2n}A_{n} \\
a_{R,2n+1}e^{i\phi }D_{n}%
\end{pmatrix} ,  \label{S1}
\\
\begin{pmatrix}
A_{n} \\
D_{n-1}%
\end{pmatrix}
= {}& S_{h} 
\begin{pmatrix}
a_{L,2n}e^{-i\phi }B_{n} \\
a_{R,2n-1}C_{n-1}%
\end{pmatrix} .  \label{S2}
\end{align}%
Solving \cref{S1,S2}, we obtain the following recurrence relations for $A_{n}$ and $B_{n}$
\begin{widetext}
\begin{gather}
	A_{1,n+1}-a_{R,2n+1}a_{L,2n}A_{1,n}=|r|a_{L,2n+2}e^{-i\phi
	}B_{1,n+1}-|r|a_{R,2n+1}B_{1,n}+a_{R,1}J_{\epsilon }^{( 1)} \delta _{n,0},   \label{Rec2}
\\
\frac{Da_{L,2n+2}a_{R,2n+1}}{1-a_{R,2n+1}^{2}e^{i\phi }}B_{1,n+1}-\left[
\frac{Da_{R,2n+1}^{2}e^{i\phi }}{1-a_{R,2n+1}^{2}e^{i\phi }}+\frac{%
	Da_{L,2n}^{2}e^{-i\phi }}{1-a_{R,2n-1}^{2}e^{i\phi }}-e^{-i\phi
}a_{L,2n}^{2}+1\right] B_{1,n}+\frac{Da_{R,2n-1}a_{L,2n}}{%
	1-a_{R,2n-1}^{2}e^{i\phi }}B_{1,n-1} \notag \\
=-|r|J_{\epsilon }^{\left( 1\right) }\delta _{n,0},  \label{Re1}
\end{gather}

Next, from the continuity equation $\frac{\partial }{\partial t}\hat{\rho}+\partial_{x}\mathcal{\hat{J}}=0$, with $\hat{\rho}=e\sum\nolimits_{\sigma }(\Psi_{\sigma ,\uparrow }^{\dag }\Psi _{\sigma, \uparrow }+\Psi _{\sigma ,\downarrow }^{\dag }\Psi _{\sigma ,\downarrow })$, we define the current operator $\mathcal{\hat{J}}=\sum\nolimits_{\sigma }\frac{ev\sigma }{\hbar }(\Psi _{\sigma ,\uparrow}^{\dag }\Psi _{\sigma ,\uparrow }-\Psi _{\sigma ,\downarrow }^{\dag }\Psi_{\sigma ,\downarrow })$. 
The average electric current in terminal 1, cf. \cref{Wave1}, is defined as 
\begin{align}
I_{1} =\langle \mathcal{\hat{J}} \rangle = {}& 
\frac{e}{h}\sum_{k,n}e^{i\frac{2keVt}{\hbar}}\int\nolimits_{-\infty}^{+\infty } \!\!d\epsilon \left[ \left( J_{\epsilon }^{\left( 1\right) } \delta _{n+k,0}+a_{L,2\left( n+k\right) }^{\ast }A_{1,n+k}^{\ast }\right)
\left( J_{\epsilon }^{\left( 1\right) }\delta _{n,0}+a_{L,2n}A_{1,n}\right)
-B_{1,n+k}^{\ast }B_{1,n}\right] f_{\epsilon }  \notag \\
&+\frac{e}{h}\sum_{k,n}e^{i\frac{2keVt}{\hbar}}\int\nolimits_{-\infty }^{+\infty} \!\! d\epsilon \left[ a_{L,2n}a_{L,2\left( n+k\right) }^{\ast} B_{1,n+k}^{\ast }B_{1,n}-A_{1,n+k}^{\ast }A_{1,n}\right] f_{-\epsilon }, 
\label{current1}
\end{align}
\end{widetext}
where $f_{\epsilon }=(e^{\epsilon /k_BT}+1)^{-1}$ is the Fermi-Dirac distribution function. 
The dc component of the current corresponds to the $k=0$ harmonic in \cref{current1}. Similarly, we can obtain the currents for the other terminals, $I_{i=2-4}$, and obtain the total current as $I=\sum_{i}I_{i}$ (see \cref{AppB}). 
In the numerical calculations, we solved \cref{Re1} by choosing an appropriate cut-off value $\left\vert n\right\vert =N$, and normalize it in units of $G_{N}\Delta /e$, where $G_{N}$ is the conductance when all electrodes are in the normal state. 
The differential conductance is thus obtained as $G=\partial I/\partial V$.

\begin{figure}[hb!]
	\begin{center}
		\includegraphics[width=85mm]{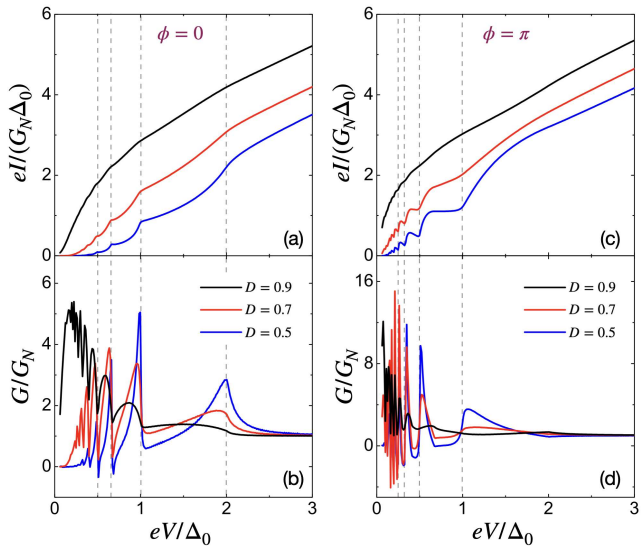}
	\end{center}
	\caption{Electric current and differential conductance at temperature $k_BT=0.2\Delta_0$ as a function of the voltage, for $\phi=0$ (a,b) and $\phi=\pi$ (c,d), for transmissions $D=0.9, 0.7, 0.5$. In all cases, $\Delta_L=\Delta_R=\Delta_0$. The dashed vertical lines indicate the positions of the resonant spikes. }
	\label{Fig3}
\end{figure}

\begin{figure}[t]
	\begin{center}
		\includegraphics[width=85mm]{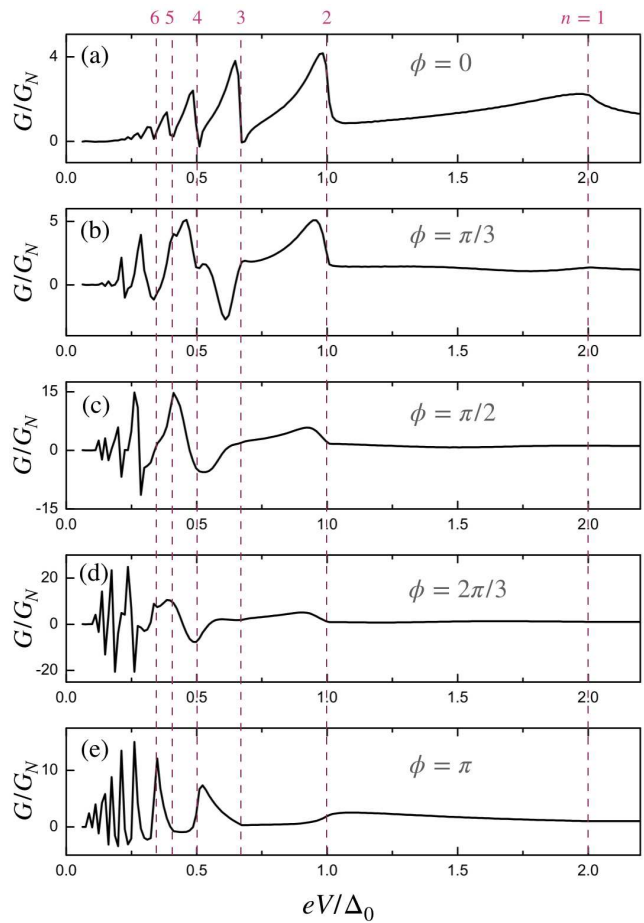}
	\end{center}
	\caption{Differential conductance as a function of the voltage at temperature $k_BT=0.2\Delta_0$ for (a) $\phi=0$, (b) $\phi=\pi/3$, (c) $\phi=\pi/2$, (d) $\phi=2\pi/3$, and (e) $\phi=\pi$. The dashed vertical lines mark the $n$-th order \gls{sgs} at $2\Delta_0/n$ for $n=1,2,3,4,5,6$. In all cases, we choose $\Delta_L=\Delta_R=\Delta_0$ and $D=0.6$. }
	\label{Fig4}
\end{figure}

\section{Subharmonic  gap structure of symmetric junctions \label{sec:SIS}}

We calculate the current and conductance in \cref{Fig3} for different junction transmissivity $D=tt^*$. We focus on the dc current which experimentally relates to the average electric current in the long time limit. Our formalism can be applied to arbitrary value of $\phi$, but, for clarity, we focus on the time-reversal invariant junction with $\phi=0$ and $\phi=\pi$. 
In \cref{Fig3}, we consider a symmetric junction with $\Delta_L=\Delta_R=\Delta_0$, and show the current and differential conductance for different values of the transmission. For $\phi=0$, the current characteristic is that of $s$-wave superconductors, where the conductance displays peaks at $eV_n=2\Delta_0/n$ ($n$ being an integer), see \cref{Fig3}(a,b). By contrast, the \gls{sgs} becomes $2\Delta_{0}/m$ ($m$ an even integer) for $\phi=\pi$, see \cref{Fig3}(c,d). 
A \gls{sgs} with only even resonances was already predicted for time-reversal breaking topological Josephson junctions with zero-energy states~\cite{Meyer11,San_Jose_2013,Zazu16}. The exotic \gls{sgs} can be understood as follows. In the topological superconducting phase, the density of states at the gap edges is suppressed, in contrast to the divergent density for trivial superconductors, see \cref{sec:NIS}. 
Thus, \gls{mar} processes where quasiparticles transmit from the lower gap edge at $-\Delta_0$ to the upper one at $\Delta_0$ will not necessarily give rise to conductance peaks. 
By contrast, when the \gls{mar} trajectory passes through the midgap \glspl{mkp}, the resonant channel will boost the \gls{mar} transmission and therefore a conductance peak appears. Consequently, the presence of zero-energy states (now \glspl{mkp}) when $\phi=\pi$ plays an important role in forming the \gls{sgs}. It is also interesting to compare our results with previous works on phase-tuned \gls{mar} in multi-terminal superconductors~\cite{Lantz02,Galaktionov12,Riwar2016,Akhmerov19}. In a conventional 3-terminal $s$-wave superconducting interferometer, the phase difference changes the visibility, instead of the shape of the \gls{sgs}. However, in our setup, the phase difference directly changes the characteristic of the \gls{sgs}.

To further show how the SGSs evolve with the phase difference, we plot the conductance spectra with equal pair potential $\Delta_L=\Delta_R=\Delta_0$ in \cref{Fig4}. Apart from the previously analyzed cases with $\phi=0$ and $\pi$, the position of the resonant peaks in the \gls{sgs} is not straightforward to identify. However, it is clear that the odd order resonances gradually disappear by increasing $\phi$ from $0$ to $\pi$. Thus, as the phase reaches $\phi=\pi$, only the even order resonant peaks remain in the \gls{sgs}.

\begin{figure*}[tb]
	\begin{center}
		\includegraphics[width=1\textwidth]{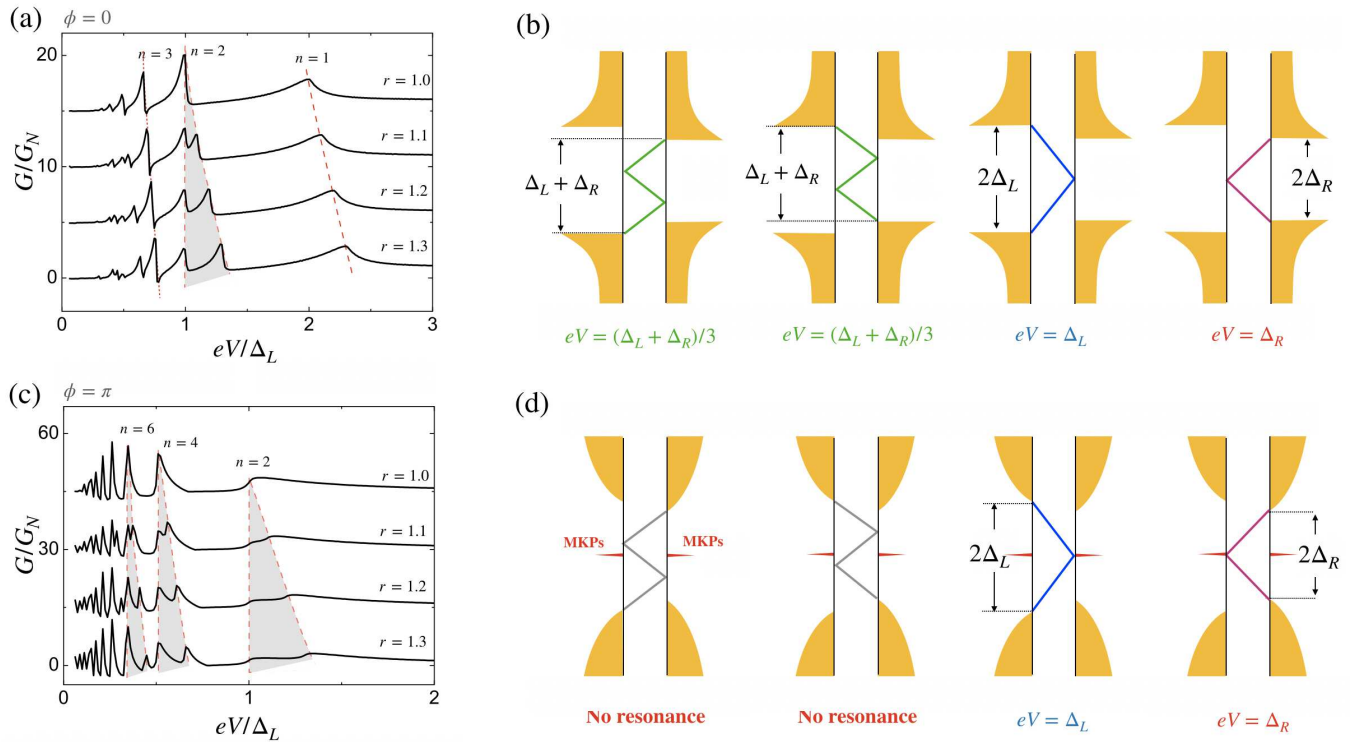}
	\end{center}
	\caption{Differential conductance at temperature $k_BT=0.2\Delta_0$ for $D=0.6$ as the gap ratio $r=\Delta_L/\Delta_R$ varies for (a) $\phi=0$ and (c) $\phi=\pi$. (b) and (d) show the schematic trajectories of the third-order (first and second panels) and the second-order (third and fourth panels) \gls{mar}. For $\phi=0$ (b), both cases result in a conductance peak, since they connect the gap edges with divergent density of states.  However, for $\phi=\pi$ (d), the density of states is reduced at the gap edges and the third-order MAR will not generate a conductance peak (gray lines). By contrast, the second-order MAR is assisted by MKPs (red and blue lines) leading to a conductance peak.} 
	\label{Fig5}
\end{figure*}

\section{Asymmetric junctions \label{sec:SIS-2}}

Next, we explore the \gls{sgs} in asymmetric junctions where $\Delta_L$ and $\Delta_R$ can be different. We focus on the time reversal invariant cases $\phi=0 ,\pi$ and compare them to the symmetric case. 
In \cref{Fig5}(a), we consider a trivial junction with $ \phi=0 $ and change the gap ratio $r=\Delta_L / \Delta_R$. 
For $\Delta_L \neq \Delta_R$, the odd order resonant conductance peaks remain at $eV_n=(\Delta_L+ \Delta_R)/n$ ($n$ odd), while the even order ones at $eV_m=2\Delta_{L/R}/m$ ($m$ even) split into two, see shaded area in \cref{Fig5}(a). 
To illustrate this difference, we sketch the {second ($n=2$) and third ($n=3$)} order \gls{mar} processes in \cref{Fig5}(b). Since the \gls{sgs} appears when \gls{mar} connect with two band edges, for odd integers (green lines) quasiparticles climb up the same energy $\Delta_L+ \Delta_R$ via \gls{mar}. Thus the \gls{sgs} for $eV_3=({\Delta_L + \Delta_R} )/3$ does not split. For even order \gls{mar}, however, the two possible paths for quasiparticles gain different energy as indicated by the blue and red lines. Therefore, the second order \gls{mar} contribute double peaks at $eV_n=\Delta_L$ and $eV_n=\Delta_R$ to the conductance spectra. 

Next, we consider the nontrivial case with $\phi=\pi$ and keep the other parameters unchanged. As for symmetric junctions, the odd order \gls{mar} resonances disappear [\cref{Fig5}(c)] due to the reduced density of states at the gap edges. 
As explained above, \gls{mar} processes connecting two gap edges only give rise to \gls{sgs} when a zero-energy state resides in its trajectory. 
It explains why the odd order gray \gls{mar} trajectories sketched in \cref{Fig5}(d) do not contribute to \gls{sgs}. 
However, the even order blue and red \gls{mar} processes satisfy the resonant condition and thus enhance the onset current leading to the appearance of conductance peaks. 
This analysis of the \gls{sgs} is also valid for time-reversal broken topological superconductors. 

It is worth highlighting that the different \glspl{sgs} between $\phi=0$ and $\pi$ are directly connected to the absence or presence, respectively, of zero-energy bound states. Since time-reversal symmetry is preserved in both cases, the emergence of zero-energy bound states should always come in degenerate pairs (\glspl{mkp}) according to Kramers theorem~\footnotemark[1]. In \cref{sec:NIS}, we test such Kramers degeneracy by proposing a different setup configuration with only one superconductor loop. There, we show the correct conductance quantization of $4e^2/h$ as corresponds to a pair of spin degenerate Majorana bound states. 

\section{Conclusions \label{sec:conc}}

We have studied the charge transport properties of quantum spin Hall strips, with coupled edge states, connected to several superconducting electrodes. 
Such a setup supports time-reversal invariant Majorana bound states, known as Majorana Kramers pairs, that appear when the phase difference at the Josephson junctions is $\phi=\pi$. 
We find that the current characteristics strongly change with the phase difference between superconductors at opposite edges. 
Consequently, the subharmonic gap structure, a sequence of resonant conductance peaks appearing in voltage-biased Josephson junctions due to multiple Andreev processes, is very sensitive to this phase difference. 
For $\phi=\pi$, due to the presence of zero-energy Majorana Kramers pairs, the odd order multiple Andreev processes do not contribute to the current, and only the even order ones appear in the subharmonic gap structure. 
Moreover, when the superconductors forming the junction have different gap sizes, all the (even) conductance peaks split, a signature without counterpart in conventional junctions. 

We now briefly discuss the feasibility of our experimental proposal to reveal time-reversal invariant Kramers pairs of Majoranas. 
The most common quantum spin Hall insulator is based on HgTe/CdTe quantum wells, where it has been reported~\cite{Hart14} that the separation between edge channels is reached for $d \sim 400$nm. The coupling strength $\alpha$ between the edge channels with this value of $d$ was estimated to be about $10\mu$eV. The superconducting gap $\Delta$ induced by proximity effect was estimated to be less than $20\mu$eV. 
In our geometry, we assume a large chemical potential $\mu$, which can be easily satisfied by tuning a top gate~\cite{Brune12}. We also consider $\Delta$ smaller than $\alpha$, which could also be realized by reducing the coupling between the superconducting leads and the quantum spin Hall edges. 
Our proposal does not require magnetic materials, thus further simplifying its experimental realization, and is highly tunable by an external magnetic flux. 
Based on these estimations and the recent advances implementing superconducting electrodes on semiconductor quantum wells~\cite{Hart14,Pribiag15,Bocquillon17,Ren2019,Fornieri2019}, we are confident that our proposal is within experimental reach. 

{\itshape Acknowledgments.} We thank Yukio Tanaka and Fanming Qu for valuable discussions. B. L. acknowledges support from the National Natural
Science Foundation of China (project 11904257) and the Natural Science Foundation of Tianjin (project 20JCQNJC01310). 
P. B. acknowledges support from the Spanish CM ``Talento Program'' project No.~2019-T1/IND-14088 and the Agencia Estatal de Investigaci\'on projects No.~PID2020-117992GA-I00 and No.~CNS2022-135950. 

\appendix

\begin{widetext}

\section{Wavefunctions}

\label{AppA} \setcounter{equation}{0} \renewcommand{\theequation}{A.%
	\arabic{equation}} In the superconducting side, we can transform the
Hamiltonian using the unitary transformation%
\begin{equation}
	U=\frac{1}{\sqrt{2}}\left[
	\begin{array}{cccc}
		1 & 1 & 0 & 0 \\
		1 & -1 & 0 & 0 \\
		0 & 0 & -1 & -1 \\
		0 & 0 & -1 & 1%
	\end{array}%
	\right] ,
\end{equation}%
and $\tilde{H}_{1\left( 2\right) }=UH_{1\left( 2\right) }U^{\dag }$ becomes%
\begin{equation}
	\tilde{H}_{1\left( 2\right) }=\left[
	\begin{array}{cccc}
		-\mu +\alpha  & \mp i\partial _{x} & \mp \gamma \xi _{c}\Delta  & \mp
		i\gamma \xi _{s}\Delta  \\
		\mp i\partial _{x} & -\mu -\alpha  & \pm i\gamma \xi _{s}\Delta  & \pm
		\gamma \xi _{c}\Delta  \\
		\mp \gamma ^{\ast }\xi _{c}\Delta  & \mp i\gamma ^{\ast }\xi _{s}\Delta  &
		\mu -\alpha  & \mp i\partial _{x} \\
		\pm i\gamma ^{\ast }\xi _{s}\Delta  & \pm \gamma ^{\ast }\xi _{c}\Delta  &
		\mp i\partial _{x} & \mu +\alpha
	\end{array}%
	\right] ,
\end{equation}%
with $\xi _{c}=\cos \frac{\phi }{2}$, $\xi _{s}=\sin \frac{\phi }{2}$, and $%
\gamma =e^{i\frac{\phi }{2}}$. It can be seen that $\alpha $ is negligible
in $\tilde{H}_{1\left( 2\right) }$ in the limit $\alpha \ll \mu $, and if $%
\phi \neq 0$ or $\pi $ there are mixed singlet- and triplet-pairings. The
wavefunction $\psi $ of $H_{1\left( 2\right) }$ can be obtained by $U^{\dag }%
\tilde{\psi}$, where $\tilde{\psi}$ is the solution of $\tilde{H}_{1\left(
	2\right) }$. The wavefunctions of $H_{1}$ in the superconducting side are%
\begin{equation}
	\psi _{1,S}^{e,\rightarrow }=\left[
	\begin{array}{c}
		u \\
		0 \\
		v \\
		0%
	\end{array}%
	\right] e^{ikx};\psi _{1,S}^{e,\leftarrow }=\left[
	\begin{array}{c}
		0 \\
		u \\
		0 \\
		ve^{-i\phi }%
	\end{array}%
	\right] e^{-ikx};\psi _{1,S}^{h,\rightarrow }=\left[
	\begin{array}{c}
		0 \\
		ve^{i\phi } \\
		0 \\
		u%
	\end{array}%
	\right] e^{-ikx};\psi _{1,S}^{h,\leftarrow }=\left[
	\begin{array}{c}
		v \\
		0 \\
		u \\
		0%
	\end{array}%
	\right] e^{ikx},
\end{equation}
and the wavefunctions of $H_{2}$ in the superconducting side are%
\begin{equation}
	\psi _{2,S}^{e,\rightarrow }=\left[
	\begin{array}{c}
		0 \\
		u \\
		0 \\
		-ve^{-i\phi }%
	\end{array}%
	\right] e^{ikx};\psi _{2,S}^{e,\leftarrow }=\left[
	\begin{array}{c}
		u \\
		0 \\
		-v \\
		0%
	\end{array}%
	\right] e^{-ikx};\psi _{2,S}^{h,\rightarrow }=\left[
	\begin{array}{c}
		-v \\
		0 \\
		u \\
		0%
	\end{array}%
	\right] e^{-ikx};\psi _{2,S}^{h,\leftarrow }=\left[
	\begin{array}{c}
		0 \\
		-ve^{i\phi } \\
		0 \\
		u%
	\end{array}%
	\right] e^{ikx},
\end{equation}
where $u$ and $v$ are the coherent factors%
\begin{equation}
	u\left( v\right) =\left[ \frac{1}{2}\pm \frac{\sqrt{\epsilon ^{2}-\Delta ^{2}%
	}}{2\epsilon }\right] ^{\frac{1}{2}}.
\end{equation}%
In the central scattering region $0<x<L$, the wavefunctions are given by%
\begin{equation}
	\psi _{1,c}^{e,1}=\left[
	\begin{array}{c}
		i \\
		1 \\
		0 \\
		0%
	\end{array}%
	\right] e^{-\kappa x};\psi _{1,c}^{e,2}=\left[
	\begin{array}{c}
		-i \\
		1 \\
		0 \\
		0%
	\end{array}%
	\right] e^{\kappa x};\psi _{1,c}^{h,1}=\left[
	\begin{array}{c}
		0 \\
		0 \\
		i \\
		1%
	\end{array}%
	\right] e^{-\kappa x};\psi _{1,c}^{h,2}=\left[
	\begin{array}{c}
		0 \\
		0 \\
		-i \\
		1%
	\end{array}%
	\right] e^{\kappa x},
\end{equation}
and%
\begin{equation}
	\psi _{2,c}^{e,1}=\left[
	\begin{array}{c}
		-i \\
		1 \\
		0 \\
		0%
	\end{array}%
	\right] e^{-\kappa x};\psi _{2,c}^{e,2}=\left[
	\begin{array}{c}
		i \\
		1 \\
		0 \\
		0%
	\end{array}%
	\right] e^{\kappa x};\psi _{2,c}^{h,1}=\left[
	\begin{array}{c}
		0 \\
		0 \\
		-i \\
		1%
	\end{array}%
	\right] e^{-\kappa x};\psi _{2,c}^{h,2}=\left[
	\begin{array}{c}
		0 \\
		0 \\
		i \\
		1%
	\end{array}%
	\right] e^{\kappa x},
\end{equation}%
with $\kappa =\alpha /v$.

\begin{figure}
	\includegraphics[width=86mm]{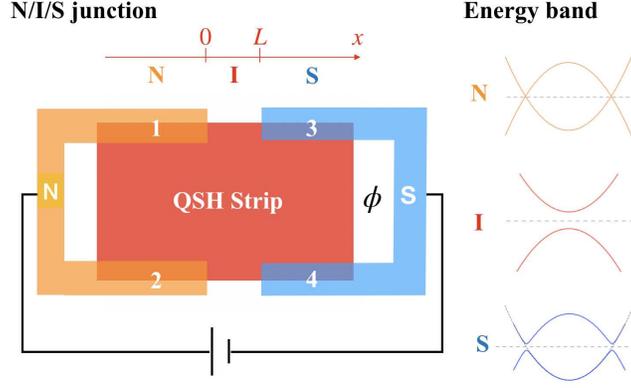}
	\caption{
		Schematic diagram of the normal-superconductor junction. Terminals 1 and 2 connect to a normal-state injection electrodes, while terminals 3 and 4 connect to superconducting leads. The superconducting phase difference between terminals is controlled by a magnetic flux. 
		Schematic band structures of low-energy quasiparticles near the Fermi surface in each region are shown on the right. }
	\label{FigS1}
\end{figure}

\section{Tunneling spectroscopy of a normal-superconductor junction \label{sec:NIS}}

 \setcounter{equation}{0} \renewcommand{\theequation}{B.%
	\arabic{equation}}

The emergence of a time-reversal invariant topological superconductor becomes apparent in the tunneling spectroscopy of a junction with superconductors present only on the right region. 
For $\phi\!=\pi $, the system preserves time-reversal symmetry and can host \glspl{mkp}. As a result, the zero-bias normal-superconductor conductance is quantized to $4e^{2}/h$. 

To show this, we now calculate the differential conductance $dI/dV$ following Ref.~\onlinecite{BTK}. 
We define the pair potential $\Delta \left( x\right) =\Delta_{0}\Theta \left( x-L\right) $ for the normal-superconductor junction, see \cref{FigS1}.  
As a result, the wavefunction for an incident electron from the normal side is
\begin{equation}
	\Phi =\sum\nolimits_{\eta =1,2}\left[ \psi _{\eta ,N}^{e,\rightarrow
	}+b_{\eta }\psi _{\eta ,N}^{e,\leftarrow }+a_{\eta }\psi _{\eta
		,N}^{h,\leftarrow }\right] .
\end{equation}

The conductance can be obtained as \cite{BTK}%
\begin{equation}
	G=G_{0}\sum\nolimits_{\eta }\left[ 1-\left\vert b_{\eta }\right\vert
	^{2}+\left\vert a_{\eta }\right\vert ^{2}\right] ,  \label{BTK}
\end{equation}%
with $G_{0}=e^{2}/h$ being the conductance quantum. 

The conductance spectra as a function of phase difference $\phi$ and the bias voltage $eV$ is shown in \cref{Fig2}(a), with transmissivity $D=tt^*=0.5$. The subgap resonance peaks vary with $\phi$ and cross at $\phi=\pi$, where the topological phase transition takes place. 
For $\phi=0$, the conductance reaches the value $G=4e^2/h$ at $eV=\pm\Delta_0$, see \cref{Fig2}(b), which indicates a perfect Andreev reflection at the gap edges~\cite{Tanaka95}. Indeed, the conductance for $\phi=0$ behaves like an $s$-wave superconductor where the subgap values reduce by decreasing the transmissivity $D$~\cite{BTK} [\cref{Fig2}(c)]. 
As expected, these quantized peaks merge at $eV=0$ for $\phi=\pi$ where the \glspl{mkp} appear, i.e., a single Majorana bound state contributes $2e^2/h$ to the conductance. Moreover, the conductance quantization remains robust against $D$ for $\phi=\pi$, exhibiting the celebrated zero-biased conductance peak due to Majorana states~\cite{Sengupta01,Bolech07,Akhmerov09,Tanaka09,Law09,Flensberg10,Crepin14,Crepin15,Ikegaya15,Ikegaya16,Bo16,Burset_2017,Keidel18,Fleckenstein18,Fleckenstein18b,Cayao_2021,Cayao_2022,Bo22} [\cref{Fig2}(d)]. At the same time, the $\pi$-difference decreases the local density of states at the gap edges $eV=\pm\Delta_0$.

\begin{figure*}
	\includegraphics[width=1\textwidth]{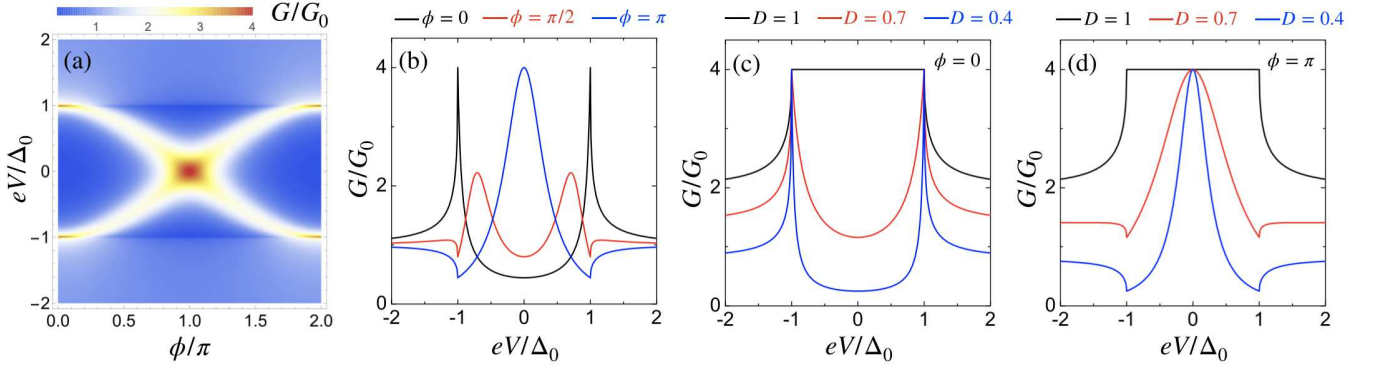}
	\caption{
		Conductance of a normal-superconductor junction: 
		(a) Contour plot of the conductance $G/G_0$, with $G_0=e^2/h$, as a function of the phase $\phi$ and bias voltage $eV$. The transmissivity is $D=0.5$. Conductance in (a) vs bias voltage for different phase differences $\phi=0$ (black), $\pi/2$ (red), and $\pi$ (blue). (c) Conductance with the phase $\phi=0$ for $D=1$ (black), $0.7$ (red) and $0.5$ (blue). 
		(d) Same as (c) for $\phi=\pi$. }
	\label{Fig2}
\end{figure*}


\section{Recursive relations and currents in the Josephson junction}\label{AppB}
\setcounter{equation}{0}
\renewcommand{\theequation}{C.\arabic{equation}}
In the main text, we have derived the current from injected quasiparticles in terminal 1. We now provide the calculation of currents induced by injection from the other three terminals. The recursive equations for a quasiparticle incident from terminal 2 are
\begin{gather}
	A_{2, n+1}-{a}_{R,2 n+1} a_{L,2 n} A_{2, n} =|r| a_{L,2 n+2} e^{i \phi} B_{2, n+1}-|r| {a}_{R,2 n+1} B_{2, n}+{a}_{R,1}J_{\epsilon }^{\left( 2\right) }\delta_{n 0} ,
\\
	\frac{D a_{L,2 n+2}{a}_{R,2 n+1} }{1-{a}_{R,2 n+1}^{2} e^{-i \phi}} B_{2, n+1}-\left[\frac{D {a}_{R,2 n+1}^2 e^{-i \phi}}{1-{a}_{R,2 n+1}^{2} e^{-i \phi}}+\frac{D a_{L,2 n}^{2} e^{i \phi}}{1-{a}_{R,2 n-1}^{2} e^{-i \phi}}-e^{i \phi }{a}_{L,2 n}^{2}+1\right] B_{2, n} +\frac{D {a}_{R,2 n-1} a_{L,2 n}}{1-{a}_{R,2 n-1}^{2} e^{-i \phi}} B_{2, n-1}
\notag \\
=-|r| J_{\epsilon }^{\left( 2\right) } \delta_{n 0}.
\end{gather}

For an incident quasiparticle from terminal 3, the relations are
\begin{gather}
	\bar{A}_{3, n+1}-\bar{a}_{L,2 n+1} {\bar{a}}_{R,2 n} \bar{A}_{3, n}=|r|{{\bar{a}}}_{R,2 n+2} e^{i \phi} \bar{B}_{3, n+1}-|r|\bar{a}_{L,2 n+1} \bar{B}_{3, n}+\bar{a}_{L,1}J_{\epsilon }^{\left( 3\right) }\delta_{n 0} ,
\\
	\frac{D \bar{a}_{L,2 n+1} {\bar{a}}_{R,2 n+2}}{1-\bar{a}_{L,2 n+1}^{2} e^{-i \phi}} \bar{B}_{3, n+1}-\left[\frac{D \bar{a}_{L,2 n+1} e^{-i \phi}}{1-\bar{a}_{L,2 n+1}^2 e^{-i \phi}}+\frac{D {\bar{a}}_{R,2 n} e^{i \phi}}{1-\bar{a}_{L,2 n-1}^2 e^{-i \phi}}-e^{i \phi} {\bar{a}}_{R,2 n}^{2}+1\right] \bar{B}_{3, n}+\frac{D \bar{a}_{L,2 n-1} {\bar{a}}_{R,2 n}}{1-\bar{a}_{L,2 n-1}^2 e^{-i \phi}} \bar{B}_{3,n-1}   \notag \\
	=-|r| J_{\epsilon }^{\left(3\right) } \delta_{n0}.
\end{gather}

And for terminal 4, they are
\begin{gather}
	\bar{A}_{4, n+1}-\bar{a}_{L,2 n+1} {\bar{a}}_{R,2 n} \bar{A}_{4, n}=|r| {\bar{a}}_{R,2 n+2} e^{-i \phi }\bar{B}_{4,n+1} -|r| \bar{a}_{L,2 n+1} \bar{B}_{4, n}+\bar{a}_{L,1} J_{\epsilon }^{\left(4\right) } \delta_{n 0} ,
\\
	\frac{D \bar{a}_{L,2 n+1}{\bar{a}}_{R,2 n+2} }{1-\bar{a}_{L,2 n+1} e^{i \phi}}\bar{B}_{4, n+1}-\left[\frac{D \bar{a}_{L,2 n+1} e^{i \phi}}{1-\bar{a}_{L,2 n+1}^{2} e^{i \phi}}+\frac{D {\bar{a}}_{R,2 n}^{2} e^{-i \phi}}{1-\bar{a}_{L,2 n-1}^{2}e^{i \phi}}-{e^{-i \phi}\bar{a}}_{R,2 n}^{2} +1\right]\bar{B}_{4, n}+\frac{D \bar{a}_{L,2 n-1} {\bar{a}}_{R,2 n}}{1-\bar{a}^2_{2 n-1}e^{i \phi}} \bar{B}_{4, n-1} \notag \\ 
	=-|r| J_{\epsilon }^{\left( 4\right) } \delta_{n0} .
\end{gather}
Here, $\bar{a}$ is defined as $\bar{a}_{L/R,n}=a_{L/R}(\epsilon-neV)$.  The current sources $J_{\epsilon }^{\left( i\right) }$ are given by $J_{\epsilon }^{\left( 2\right) }=%
\sqrt{1-\left\vert a_{L}\left( \epsilon \right) \right\vert ^{2}}$ and $%
J_{\epsilon }^{\left( 3\right) }=J_{\epsilon }^{\left( 4\right) }=\sqrt{%
	1-\left\vert a_{R}\left( \epsilon \right) \right\vert ^{2}}$. We have
defined ${A}_{n}=A(n,eV)$, ${B}_{n}=B(n,eV)$, $\bar{A}_{n}=A(n,-eV)$ and $%
\bar{B}_{n}=B(n,-eV)$. The resulting currents $I_2$, $I_3$, and $I_4$ are derived as
\begin{align}
	I_{2} ={}&\frac{e}{h}\sum\nolimits_{k}e^{i\frac{2keVt}{\hbar}}\int\nolimits_{-\infty
	}^{+\infty }d\epsilon \sum_{n}\left[ \left( J_{\epsilon }^{\left( 2\right)
	}\delta _{n+k,0}+a_{L,2\left( n+k\right) }^{\ast }A_{2,n+k}^{\ast }\right)
	\left( J_{\epsilon }^{\left( 2\right) }\delta _{n,0}+a_{L,2n}A_{2,n}\right)
	-B_{2,n+k}^{\ast }B_{2,n}\right] f_{\epsilon }  \nonumber \\
	&+\frac{e}{h}\sum\nolimits_{k}e^{i\frac{2keVt}{\hbar}}\int\nolimits_{-\infty }^{+\infty
	}d\epsilon \sum_{n}\left[ a_{L,2n}a_{L,2\left( n+k\right) }^{\ast
	}B_{2,n+k}^{\ast }B_{2,n}-A_{2,n+k}^{\ast }A_{2,n}\right] f_{-\epsilon },
	\label{current2}
\\
	I_{3} ={}&\frac{e}{h}\sum\nolimits_{k}e^{i\frac{2keVt}{\hbar}}\int\nolimits_{-\infty
	}^{+\infty }d\epsilon \sum_{n}\left[- \left( J_{\epsilon }^{\left( 3\right)
	}\delta _{n+k,0}+\bar{a}_{R,2\left( n+k\right) }^{\ast }\bar{A}_{3,n+k}^{\ast }\right)
	\left( J_{\epsilon }^{\left( 3\right) }\delta _{n,0}+\bar{a}_{R,2n}\bar{A}_{3,n}\right)
	+\bar{B}_{3,n+k}^{\ast }\bar{B}_{3,n}\right] f_{\epsilon }  \nonumber \\
	&+\frac{e}{h}\sum\nolimits_{k}e^{i\frac{2keVt}{\hbar}}\int\nolimits_{-\infty }^{+\infty
	}d\epsilon \sum_{n}\left[- \bar{a}_{R,2n}a_{R,2\left( n+k\right) }^{\ast
	}\bar{B}_{3,n+k}^{\ast }\bar{B}_{3,n}+\bar{A}_{3,n+k}^{\ast }\bar{A}_{3,n}\right] f_{-\epsilon },
	\label{current3}
\\
I_{4} ={}&\frac{e}{h}\sum\nolimits_{k}e^{i\frac{2keVt}{\hbar}}\int\nolimits_{-\infty
}^{+\infty }d\epsilon \sum_{n}\left[- \left( J_{\epsilon }^{\left( 4\right)
}\delta _{n+k,0}+\bar{a}_{R,2\left( n+k\right) }^{\ast }\bar{A}_{4,n+k}^{\ast }\right)
\left( J_{\epsilon }^{\left( 4\right) }\delta _{n,0}+\bar{a}_{R,2n}\bar{A}_{4,n}\right)
+\bar{B}_{4,n+k}^{\ast }\bar{B}_{4,n}\right] f_{\epsilon }  \nonumber \\
&+\frac{e}{h}\sum\nolimits_{k}e^{i\frac{2keVt}{\hbar}}\int\nolimits_{-\infty }^{+\infty
}d\epsilon \sum_{n}\left[- \bar{a}_{R,2n}a_{R,2\left( n+k\right) }^{\ast
}\bar{B}_{4,n+k}^{\ast }\bar{B}_{4,n}+\bar{A}_{4,n+k}^{\ast }\bar{A}_{4,n}\right] f_{-\epsilon }.
\label{current4}
\end{align}
\end{widetext}


\bibliography{josephson}
\end{document}